# A Verifiable Fully Homomorphic Encryption Scheme for Cloud Computing Security

Ahmed EL-YAHYAOUI

Information security research team, CEDOC ST2I, ENSIAS

Mohammed V University in Rabat
, Morocco

Ahmed_elyahyaoui@um5.ac.ma

Mohamed Dafir ECH-CHRIF EL KETTANI

Information security research team, CEDOC ST2I, ENSIAS

Mohammed V University in Rabat

Rabat, Morocco

Dafir.elkettani@um5.ac.ma

*Abstract—*: Performing smart computations in a context of cloud computing and big data is highly appreciated today. Fully homomorphic encryption (FHE) is a smart category of encryption schemes that allows working with the data in its encrypted form. It permits us to preserve confidentiality of our sensible data and to benefit from cloud computing powers. Currently, it has been demonstrated by many existing schemes that the theory is feasible but the efficiency needs to be dramatically improved in order to make it usable for real applications. One subtle difficulty is how to efficiently handle the noise. This paper aims to introduce an efficient and verifiable FHE based on a new mathematic structure that is noise free.

## I. INTRODUCTION

Cloud computing has manifested as a powerful computing model in the last decade, with numerous advantages both to clients and providers. One of the obvious huge advantage is that clients can delegate their complex computations and benefit from the best technologies and computation powers at low costs. The cost benefits presented by cloud technologies is one of the major arguments that justify the spreading of cloud computing in many industries. During the last few years, enterprises culture of accepting cloud computing was developed and many companies had shown their ready to adhere cloud and benefit from its capacities, but businesses are now finding that there is a number of security issues that have to be treated when venturing into the cloud.

Privacy of sensible data is one of most important security issues. Leakage of some data can cause huge damages to its owners. In general, to save privacy of our data it is advised to encrypt it before storing it on a remote cloud server. Using classical encryption schemes as RSA, AES, 3DES… allows clients to preserve data privacy during transmission to the cloud, but if a client requests the cloud to perform a complex treatment on its data, he should share his private key with the remote cloud server. This traditional use of cryptography may not be the best solution in terms of privacy, especially if we consider the cloud as an untrusted part.

One solution to this problematic is doing smart computations on encrypted data, this idea was early introduced by Rivest, Adleman and Dertozous in 1978[1], authors conjectured the existence of a privacy homomorphism. Today we are using the notion of Fully Homomorphic Encryption (FHE) rather than privacy homomorphism.

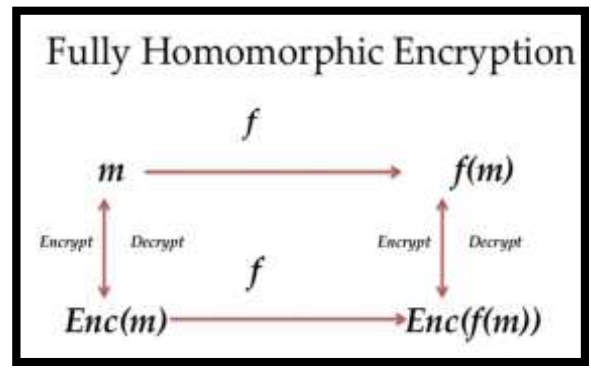

*Figure 1: Fully Homomorphic Encryption diagram*

FHE schemes (figure 1) are considered as the next generation algorithms for cryptography, it is a type of smart encryption cryptosystems that support arbitrary computations on ciphertexts without ever needing to decrypt or reveal it. In a context of cloud computing and distributed computation this is a highly precious power. In fact, a significant application of fully homomorphic encryption is to big data and cloud computing. Generally, FHE is used in outsourcing complex computations on sensitive data stored in a cloud as it can be employed in specific applications for big data like secure search on encrypted big data and private information retrieval. It was an open problem until the revolutionary work of Gentry in 2009 [2]. In his thesis, Gentry proposed the first adequate fully homomorphic encryption scheme by exploiting properties of ideal lattices.

Gentry's construction is based on his bootstrapping theorem which provides that given a somewhat homomorphic encryption scheme (SWHE) that can evaluate homomorphically its own decryption circuit and an additional NAND gate, we can pass to a "levelled" fully homomorphic encryption scheme and so obtain a FHE scheme by assuming circular security. The purpose of using bootstrapping technique is to allow refreshment of ciphertexts and reduce noise after its growth.

Gentry's construction is not a single algorithm but it considered as a framework that inspires cryptologists to build new fully homomorphic encryption schemes [3, 4, 5, 6…]. A FHE cryptosystem that uses Gentry's bootstrapping technique can be classified in the category of noise-based fully homomorphic encryption schemes [7]. If this class of cryptosystems has the advantage to be robust and more secure, it has the drawback to be not efficient in terms of runtime and ciphertext size. In several works followed Gentry's one, many techniques of noise management are invented to improve runtime efficiency and to minimise ciphertext and key size's [8,9,10...], but the problematic of designing a practical and efficient fully homomorphic encryption scheme remains the same until now.

In the literature we can locate a second category called free-noise fully homomorphic encryption schemes which do not need a technique of noise management to refresh ciphertexts. In a free-noise fully homomorphic encryption scheme one can do infinity of operations on the same ciphertext without noise growing. This class of encryption schemes is known as faster than the previous one, involves simple operations to evaluate circuits on ciphertexts and do not require a noise management technique, but it suffers from security problems because the majority of designed schemes are cryptanalyzed today.

A verifiable encryption scheme is a cryptosystem that allows us to prove some properties about an encrypted value without disclosing it. If the verification option is combined with homomorphic capacities in the same encryption scheme, it becomes a verifiable fully homomorphic encryption scheme (VFHE). Consequently, a VFHE scheme (figure 2) is a very smart scheme that we can use to outsource complex computations on sensible data to a remote cloud server. It allows the client to verify the correctness of its delegated computations.

In this work, we will adopt the free-noise approach to design an efficient verifiable fully homomorphic encryption scheme. We will try to overcome the problem of weak security through using the ring of quaternions.

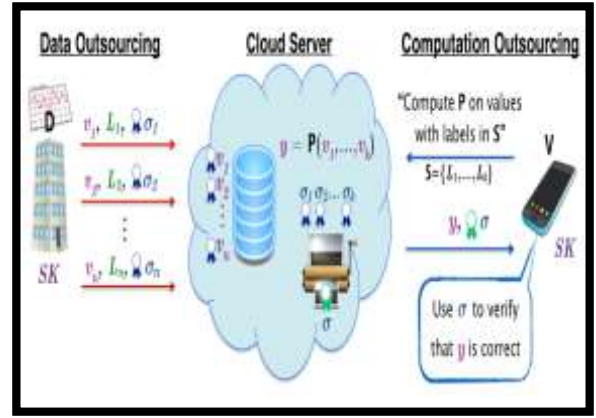

*Figure 2: Verifiable Fully Homomorphic Encryption diagram*

## II. OUR TECHNIQUES AND RESULTS

We propose an efficient and verifiable noise-free fully homomorphic encryption scheme that uses the ring of Lipschitz's quaternions and permits computations over encrypted data under a symmetric key; our scheme permits us to verify if the computation was performed in its correct form. We exploit properties of non-commutativity of Lipschitz integers to build our efficient encryption scheme.

## III. MATHEMATICAL BACKGROUND
### A. Quaternionique field $\mathbb{H}$

A quaternion is a number in a general sense. Quaternions encompass real and complex numbers in a number system where multiplication is no longer a commutative law.

The quaternions were introduced by the Irish mathematician William Rowan Hamilton in 1843. They now find applications in mathematics, physics, computer science and engineering.

Mathematically, the set of quaternions $\mathbb{H}$ is a non-commutative associative algebra on the field of real numbers $\mathbb{R}$ generated by three elements $i, j$ and $k$ satisfying relations: $i^2 = j^2 = k^2 = i.j.k = -1$. Concretely, any quaternion $q$ is written uniquely in the form: $q = a + bi + cj + dk$ where $a, b, c$ and $d$ are real numbers.

The operations of addition and multiplication by a real scalar are trivially done term to term, whereas the multiplication between two quaternions is termed by respecting the non-commutativity and the rules proper to $i, j$ and $k$. For example, given $q = a + bi + cj + dk$ and $q' = a' + b'i + c'j + d'k$ we have $qq' = a_0 + b_0 i + c_0 j + d_0 k$ such that: $a_0 = aa' - (bb' + cc' + dd')$, $b_0 = ab' + a'b + cd' - c'd$, $c_0 = ac' - bd' + ca' + db'$ and $d_0 = ad' + bc' - cb' + a'd$.

The quaternion $\bar{q} = a - bi - cj - dk$ is the conjugate of $q$.

$|q| = \sqrt{q\bar{q}} = \sqrt{a^2 + b^2 + c^2 + d^2}$ is the module of $q$. The real part of $q$ is $Re(q) = \frac{q+\bar{q}}{2} = a$ and the imaginary part is $Im(q) = \frac{q-\bar{q}}{2} = bi + cj + dk$.

A quaternion $q$ is invertible if and only if its modulus is non-zero, and we have $q^{-1} = \frac{1}{|q|^2}\bar{q}$.

### B. Reduced form of quaternion:

Quaternion can be represented in a more economical way, which considerably alleviates the calculations and highlights interesting results. Indeed, it is easy to see that $\mathbb{H}$ is a $\mathbb{R}$-vectorial space of dimension 4, of which $(1, i, j, k)$ constitutes a direct orthonormal basis. We can thus separate the real component of the pure components, and we have for $q \in \mathbb{H}, q = (a, \mathbf{u})$ such that $\mathbf{u}$ is a vector of $\mathbb{R}^3$. So for $q = (a, \mathbf{u}), q' = (a', \mathbf{v}) \in \mathbb{H}$ and $\lambda \in \mathbb{R}$ we obtain:

1. $q + q' = (a + a', \mathbf{u} + \mathbf{v})$ and $\lambda q = (\lambda a, \lambda \mathbf{u})$
2. $qq' = (aa' - \mathbf{u}.\mathbf{v}, a\mathbf{v} + a'\mathbf{u} + \mathbf{u} \wedge \mathbf{v})$ Where $\wedge$ is the cross product of $\mathbb{R}^3$.
3. $\bar{q} = (a, -\mathbf{u})$ and $|q|^2 = a^2 + \mathbf{u}^2$.

### C. Ring of Lipschitz integers

The set of quaternions defined as follows:

$\mathbb{H}(\mathbb{Z}) = \{q = a + bi + cj + dk / a, b, c, d \in \mathbb{Z}\}$ has a ring structure called the ring of Lipschitz integers, $\mathbb{H}(\mathbb{Z})$ is trivially non-commutative.

For r $n \in \mathbb{N}^*$, the set of quaternions:

$\mathbb{H}(\mathbb{Z}/n\mathbb{Z}) = \{q = a + bi + cj + dk / a, b, c, d \in \mathbb{Z}/n\mathbb{Z}\}$ has the structure of a non-commutative ring.

A modular quaternion of Lipschitz q $\in \mathbb{H}(\mathbb{Z}/n\mathbb{Z})$ is invertible if and only if its module and the integer $\mathbf{n}$ are coprime numbers, i.e $|q|^2 \wedge n = 1$.

### D. Quaternionique matrices $\mathbb{M}_2(\mathbb{H}(\mathbb{Z}/n\mathbb{Z}))$:

The set of matrices $\mathbb{M}_2(\mathbb{H}(\mathbb{Z}/n\mathbb{Z}))$ describes the matrices with four inputs (two rows and two columns) which are quaternions of $\mathbb{H}(\mathbb{Z}/n\mathbb{Z})$. This set has a non-commutative ring structure.

There are two ways of multiplying the quaternion matrices: The Hamiltonian product, which respects the order of the factors, and the octonionique product, which does not respect it.

The Hamiltonian product is defined as for all matrices with coefficients in a ring (not necessarily commutative). For example:

$$U = \begin{pmatrix} u_{11} & u_{12} \\ u_{21} & u_{22} \end{pmatrix}, \quad V = \begin{pmatrix} v_{11} & v_{12} \\ v_{21} & v_{22} \end{pmatrix}$$

$$\Rightarrow UV = \begin{pmatrix} u_{11}v_{11} + u_{12}v_{21} & u_{11}v_{12} + u_{12}v_{22} \\ u_{21}v_{11} + u_{22}v_{21} & u_{21}v_{12} + u_{22}v_{22} \end{pmatrix}$$

The octonionique product does not respect the order of the factors: on the main diagonal, there is commutativity of the second products and on the second diagonal there is commutativity of the first products.

$$U = \begin{pmatrix} u_{11} & u_{12} \\ u_{21} & u_{22} \end{pmatrix}, \quad V = \begin{pmatrix} v_{11} & v_{12} \\ v_{21} & v_{22} \end{pmatrix}$$

$$\Rightarrow UV = \begin{pmatrix} u_{11}v_{11} + v_{21}u_{12} & v_{12}u_{11} + u_{12}v_{22} \\ v_{11}u_{21} + u_{22}v_{21} & u_{21}v_{12} + v_{22}u_{22} \end{pmatrix}$$

In our article we will adopt the Hamiltonian product as an operation of multiplication of the quaternionique matrices.

### E. Schur complement and inversibility of quaternionique matrices

Let $\mathcal{R}$ be an arbitrary associative ring, a matrix $M \in \mathcal{R}^{n \times n}$ is supposed to be invertible if $\exists N \in \mathcal{R}^{n \times n}$ such that $MN = NM = I_n$ where $N$ is necessarily unique.

The Schur complement method is a very powerful tool for calculating inverse of matrices in rings. Let $M \in \mathcal{R}^{n \times n}$ be a matrix per block satisfying:

$$M = \begin{pmatrix} A & B \\ C & D \end{pmatrix} \text{ such that } A \in \mathcal{R}^{k \times k}.$$

Suppose that $A$ is invertible, we have:

$$M = \begin{pmatrix} I_k & 0 \\ CA^{-1} & I_{n-k} \end{pmatrix} \begin{pmatrix} A & 0 \\ 0 & A_s \end{pmatrix} \begin{pmatrix} I_k & A^{-1}B \\ 0 & I_{n-k} \end{pmatrix} \text{ where}$$

$A_s = D - CA^{-1}B$ is the Schur complement of $A$ in $M$.

The inversibility of $A$ ensures that the matrix $M$ is invertible if and only if $A_s$ is invertible. The inverse of $M$ is:

$$M^{-1} = \begin{pmatrix} I_k & -A^{-1}B \\ 0 & I_{n-k} \end{pmatrix} \begin{pmatrix} A^{-1} & 0 \\ 0 & A_s^{-1} \end{pmatrix} \begin{pmatrix} I_k & 0 \\ -CA^{-1} & I_{n-k} \end{pmatrix}$$

$$= \begin{pmatrix} A^{-1} + A^{-1}BA_s^{-1}CA^{-1} & -A^{-1}BA_s^{-1} \\ -A_s^{-1}CA^{-1} & A_s^{-1} \end{pmatrix}.$$

For a quaternionique matrix

$M = \begin{pmatrix} a & b \\ c & d \end{pmatrix} \in \mathcal{R}^{2\times 2} = \mathbb{M}_2(\mathbb{H}(\mathbb{Z}/n\mathbb{Z}))$ where the quaternion $a$ is invertible as well as its Schur complement $a_s = d - ca^{-1}b$ we have $M$ is invertible and: $M^{-1} = \begin{pmatrix} a^{-1} + a^{-1}ba_s^{-1}ca^{-1} & -a^{-1}ba_s^{-1} \\ -a_s^{-1}ca^{-1} & a_s^{-1} \end{pmatrix}$.

Therefore, to randomly generate an invertible quaternionique matrix, it suffices to:

- Choose randomly three quaternions $a, b\ and\ c$ for which $a$ is invertible.
- Select randomly the fourth quaternion $d$ such that the Schur complement $a_s = d - ca^{-1}b$ of $a$ in $M$ is invertible.

## IV. A VERIFIABLE FHE SCHEME

We place ourselves in a context where Bob wants to store confidential data in a very powerful but non-confident cloud. Bob will later need to execute complex processing on his data, of which he does not have the necessary computing powers to perform it. At this level he thinks for, at first, the encryption of his sensitive data to avoid any fraudulent action. But the ordinary encryption, which he knows, does not allow the cloud to process his calculation requests without having decrypted the data stored beforehand, which impairs their confidentiality. Bob asks if there is a convenient and efficient type of encryption to process his data without revealing it to the cloud. The answer to Bob's question is favorable, in fact since 2009 there exist so-called fully homomorphic encryption, the principle of which is quite simple: doing computations on encrypted data without thinking of any previous decryption.

As the cloud is unconfident, computations over encrypted data may be false or done incorrectly. Bob must have tools to verify the veracity of the demanded computations. For this purpose, the cloud must show Bob a proof, which can be verified by Bob on receipt, of the exactitude of the performed operations. This proof is an additional service offered by the used fully homomorphic scheme.

In order to profitably benefit from the technological advance of the cloud computing and to outsource its heavy calculations comfortably, Bob needs a robust highly secure and verifiable fully homomorphic encryption scheme whose operations, addition and multiplication, are done in a judicious time, whose noise generated during a treatment is manageable and of which he has a proof of exactitude of the performed operations on encrypted data.

To help Bob take full advantage of the powers of the cloud, we introduce a probabilistic symmetric and verifiable fully homomorphic encryption scheme without noise. The addition and multiplication operations generate no noise. We can describe our cryptosystem as follows:

### *Key generation*

-Bob generates randomly two big prime numbers p and q.

-Then, he calculates $N = p.q$.

-Bob generates randomly an invertible matrix

$$K = \begin{pmatrix} k_1 & k_2 \\ k_3 & k_4 \end{pmatrix} = \begin{pmatrix} a_{11} & a_{12} & a_{13} & a_{14} \\ a_{21} & a_{22} & a_{23} & a_{24} \\ a_{31} & a_{32} & a_{33} & a_{34} \\ a_{41} & a_{42} & a_{43} & a_{44} \end{pmatrix} \in \mathbb{M}_4(\mathbb{H}(\mathbb{Z}/N^2\mathbb{Z}))$$

such that $k_1$ is an invertible matrix.

- Bob calculates the inverse of $K$ and $k_1$, Which will be denoted $K^{-1}$ and $k_1^{-1}$.

-The secrete key is $(k_1, k_1^{-1}, K, K^{-1})$.

### *Encryption*

Lets $\sigma \in \mathbb{Z}/N^2\mathbb{Z}$ be a clear text. To encrypt $\sigma$ Bob proceed as follows:

- Bob transforms $\sigma$ into a quaternion:

$m = \sigma + \alpha Ni + \beta Nj + \gamma Nk \in \mathbb{H}(\mathbb{Z}/N^2\mathbb{Z})$ such that $\alpha, \beta, \gamma \in \mathbb{Z}/N\mathbb{Z}$.

-Bob generates a matrix: $M = \begin{pmatrix} m & r_1 \\ 0 & r_2 \end{pmatrix} \in \mathbb{M}_2(\mathbb{H}(\mathbb{Z}/N^2\mathbb{Z}))$ such that $r_1$ and $r_2 \in \mathbb{H}(\mathbb{Z}/N^2\mathbb{Z})$ are randomly generated.

-Bob calculates $M' = k_1 M k_1^{-1}$.

- Bob transforms $\sigma$ into a quaternion:

$m' = \sigma + \alpha' Ni + \beta' Nj + \gamma' Nk \in \mathbb{H}(\mathbb{Z}/N^2\mathbb{Z})$.

-Bob generates a matrix $M'' = \begin{pmatrix} m' & r_1' \\ 0 & 0 \end{pmatrix} \in \mathbb{M}_2(\mathbb{H}(\mathbb{Z}/N^2\mathbb{Z}))$ such that $r_1' \in \mathbb{H}(\mathbb{Z}/N^2\mathbb{Z})$

-The ciphertext of $\sigma$ is:

$C = Enc(\sigma) = K \begin{pmatrix} M' & R \\ 0 & M'' \end{pmatrix} K^{-1} \in \mathbb{M}_4(\mathbb{H}(\mathbb{Z}/N^2\mathbb{Z}))$ such that $R \in \mathbb{M}_2(\mathbb{H}(\mathbb{Z}/N^2\mathbb{Z}))$ is randomly generated.

### *Decryption and verification*

Lets $C \in \mathbb{M}_4(\mathbb{H}(\mathbb{Z}/N^2\mathbb{Z}))$ be a ciphertext. To decrypt $C$, using his secret key $(k_1, k_1^{-1}, K, K^{-1})$, Bob proceed as follows:

-He calculates $\begin{pmatrix} M' & R \\ 0 & M'' \end{pmatrix} = K^{-1}CK$.

-He computes $M = k_1^{-1} M' k_1$.

- Then he takes the first inputs of the resulting matrices $m = (M)_{1,1}$ and $m' = (M')_{1,1}$

- Finally, he recovers his clear message by verifying if $\sigma = m \bmod N = m' \bmod N$. If the verification is true, then the clear message returned is σ otherwise the ciphertext has been modified.

### *Addition and multiplication:*

Let $\sigma_1$ and $\sigma_2$ be two clear texts and $C_1 = Enc(\sigma_1)$ and $C_2 = Enc(\sigma_2)$ be their ciphertexts respectively.

It is easy to verify that:

(1) $C_{mult} = C_1 \times C_2 \bmod N^2 = Enc(\sigma_1) \times Enc(\sigma_2) \bmod N^2 = Enc(\sigma_1 \times \sigma_2)$.
(2) $C_{add} = C_1 + C_2 \bmod N^2 = Enc(\sigma_1) + Enc(\sigma_2) \bmod N^2 = Enc(\sigma_1 + \sigma_2)$.

## V. Security of the Scheme

Ciphertext indistinguishability is an important security property of many encryption schemes. Intuitively, if a cryptosystem possesses the property of indistinguishability, then an adversary will be unable to distinguish pairs of ciphertexts based on the message they encrypt. It is easy to see that a fully homomorphic encryption scheme cannot be secure against adaptive chosen ciphertext attacks (IND-CCA2).

### *The adversary:*

We are protecting ourselves from an adversary $\mathcal{A}$, who:

- Is a probabilistic polynomial time Turing machine.
- Has all the algorithms.
- Has full access to communication media.

### *Chosen Ciphertext Attack*

In this model, the attack assumes that the adversary $\mathcal{A}$ has access to an encryption oracle and that the adversary can choose an arbitrary number of plaintexts to be encrypted and obtain the corresponding ciphertexts. In addition, the adversary $\mathcal{A}$ gains access to a decryption oracle, which decrypts arbitrary ciphertexts at the adversary's request, returning the plaintext.

### *Startup*

1. The challenger generates a secret key $Sk$ based on some security parameter $k$ (e.g., a key size in bits) and retains it.
2. The adversary $\mathcal{A}$ may ask the encryption oracle for any number of encryptions, calls to the decryption oracle based on arbitrary ciphertexts, or other operations.
3. Eventually, the adversary $\mathcal{A}$ submits two distinct chosen plaintexts $m_0, m_1$ to the challenger.

### *The Challenge*

4. The challenger selects a bit $b \in \{0,1\}$ uniformly at random, and sends the "challenge" ciphertext $C = Enc(Sk, m_b)$ back to the adversary. The adversary is free to perform any number of additional computations or encryptions.
   a. In the non-adaptive case (IND-CCA), the adversary may not make further calls to the decryption oracle before guessing.
   b. In the adaptive case (IND-CCA2), the adversary may make further calls to the decryption oracle, but may not submit the challenge ciphertext C.
5. In the end it will guess the value of $b$.

### *The Result*

- Again, the adversary $\mathcal{A}$ wins the game if it guesses the bit $b$.
- A cryptosystem is indistinguishable under chosen ciphertext attack if no adversary can win the above game with probability $p$ greater than $\frac{1}{2} + \varepsilon$ where is a negligible function in the security parameter $k$.
- If $p > \frac{1}{2}$ then the difference $p - \frac{1}{2}$ is the advantage of the given adversary in distinguishing the ciphertext.

In our situation, the adversary $\mathcal{A}$ should distinguish an encryption of zero from an encryption of one after asking the encryption oracle of a number of encryptions and the decryption oracle to decrypt arbitrary ciphertexts. The adversary $\mathcal{A}$ can do operations on the two given ciphertexts to distinguish zero from one, as he can do operations on the entire ciphertext matrices or just to use some entrees (the diagonal of ciphertexts matrices). In our case, even if the diagonal of $M$ determines completely the invertibility of $C$, an encryption of a cleartext $\sigma \in \mathbb{Z}/N^2\mathbb{Z}$ is always non invertible because of the choice of the last component of the matrix $M''$, which is null. Therefore, an adversary cannot then distinguish encryptions of units from encryptions of non-units. Consequently, the attack proposed on Li-Wang's scheme [10] in [12] do not work for our case. Based on these assumptions, we believe that our fully homomorphic encryption scheme is indistinguishable under chosen ciphertext attacks (IND-CCA1).

Concerning the security of the secret key:

Given a random secret key of our encryption scheme: $K = \begin{pmatrix} a_{11} & a_{12} & a_{13} & a_{14} \\ a_{21} & a_{22} & a_{23} & a_{24} \\ a_{31} & a_{32} & a_{33} & a_{34} \\ a_{41} & a_{42} & a_{43} & a_{44} \end{pmatrix}$ and $K^{-1} = \begin{pmatrix} \overline{a_{11}} & \overline{a_{12}} & \overline{a_{13}} & \overline{a_{14}} \\ \overline{a_{21}} & \overline{a_{22}} & \overline{a_{23}} & \overline{a_{24}} \\ \overline{a_{31}} & \overline{a_{32}} & \overline{a_{33}} & \overline{a_{34}} \\ \overline{a_{41}} & \overline{a_{42}} & \overline{a_{43}} & \overline{a_{44}} \end{pmatrix}$. And cleartext $\sigma \in \mathbb{Z}/N^2\mathbb{Z}$.

A ciphertext of $m = \sigma + \alpha N i + \beta N j + \gamma N k \in \mathbb{H}(\mathbb{Z}/N^2\mathbb{Z})$ such that $\alpha, \beta, \gamma \in \mathbb{Z}/N\mathbb{Z}$, is determined by: $C = K \begin{pmatrix} M' & R \\ 0 & M'' \end{pmatrix} K^{-1} = \begin{pmatrix} c_{11} & c_{12} & c_{13} & c_{14} \\ c_{21} & c_{22} & c_{23} & c_{24} \\ c_{31} & c_{32} & c_{33} & c_{34} \\ c_{41} & c_{42} & c_{43} & c_{44} \end{pmatrix}$ such that $M = \begin{pmatrix} m & r_3 & r_4 \\ 0 & r_1 & r_5 \\ 0 & 0 & r_2 \end{pmatrix}$.

Therefore, we obtain sixteen equations issued from matrix operations.

According to the decryption algorithm, the plaintext $m$ can be obtained by the equation:

$(1)\ m = (\overline{a_{1,1}}c_{1,1} + \overline{a_{1,2}}c_{2,1} + \overline{a_{1,3}}c_{3,1} + \overline{a_{1,4}}c_{4,1})a_{1,1} + (\overline{a_{1,1}}c_{1,2} + \overline{a_{1,2}}c_{2,2} + \overline{a_{1,3}}c_{3,2} + \overline{a_{1,4}}c_{4,2})a_{2,1} + (\overline{a_{1,1}}c_{1,3} + \overline{a_{1,2}}c_{2,3} + \overline{a_{1,3}}c_{3,3} + \overline{a_{1,4}}c_{4,3})a_{3,1} + (\overline{a_{1,1}}c_{1,4} + \overline{a_{1,2}}c_{2,4} + \overline{a_{1,3}}c_{3,4} + \overline{a_{1,4}}c_{4,4})a_{4,1}$

An adversary who possesses the ciphertext $C$ and wants to find the cleartext m or the secret key from the above sixteen equations should, at least, extract the secret components $\overline{a_{1,1}}$, $\overline{a_{1,2}}$, $\overline{a_{1,3}}$, $\overline{a_{1,4}}$ $a_{1,1}$, $a_{2,1}$, $a_{3,1}$ and $a_{4,1}$ according to the equation (1). Since our fully homomorphic encryption scheme is probabilistic, these sixteen equations are randomly independent even if the encrypted messages are the same one. Therefore finding the secret key is equivalent to a problem of solving an over-defined system of quadratic multivariate polynomial equations in a non-commutative ring.

## VI. Comparison with other schemes

*Table 1: comparison of the performances of FHE schemes*

| Algorithm | Cleartext space | Secret key size | Public key size | Ciphertext size |
|---|---|---|---|---|
| Gentry[2] | {0,1} | $n^7$ | $n^3$ | $n^{1.5}$ |
| Smart-Verc[3] | {0,1} | $O(n^3)$ | $n^3$ | $O(n^{1.5})$ |
| DGHV[4] | {0,1} | $\tilde{O}(\lambda^{10})$ | $\tilde{O}(\lambda^2)$ | $\tilde{O}(\lambda^5)$ |
| CMNT[8] | {0,1} | $\tilde{O}(\lambda^7)$ | $\tilde{O}(\lambda^2)$ | $\tilde{O}(\lambda^5)$ |
| Batch DGHV[9] | $\{0,1\}^l$ | $\tilde{O}(\lambda^7)$ | $l.\tilde{O}(\lambda^2)$ | $l.\tilde{O}(\lambda^5)$ |
| LI-WANG[10] | $\mathbb{Z}/N\mathbb{Z}$ | $O(3N)$ | NA | $O(3N)$ |
| Our scheme | $\mathbb{Z}/N^2\mathbb{Z}$ | $O(16N^2)$ | NA | $O(16N^2)$ |

As it is shown in table 1, our cryptosystem presents good performances compared to other existing schemes. Its ciphertext and key sizes depend linearly to cleartext space dimension. The other schemes use a small cleartext space which influences the runtime of the algorithm. In our case we are using a large cleartext space which allows us to encrypt big messages and perform computations directly on ciphertexts. We can observe that the complexity of Li-Wang's scheme is smaller than ours, but this scheme uses a smaller cleartext space and it is not a verifiable scheme.

## VII. CONCLUSION AND PERSPECTIVES:

In this paper, we presented a new verifiable fully homomorphic encryption scheme. It is symmetric, noise free and probabilistic cryptosystem, for which the ciphertext space is a non-commutative ring quaternionic based. The cleartext can be a large number $\sigma \in \mathbb{Z}/N^2\mathbb{Z}$. Our encryption scheme finds its effective applications in the domain of smart computations on encrypted data in cloud computing as it can be applied also to big data security. It is an efficient and practical scheme whose security is based on the problem of solving an over-defined system of quadratic multivariate polynomial equations in a non-commutative ring. In the next work we will implement this cryptosystem and proof its security.

### REFERENCES

[1] R. Rivest, . L. Adleman and M. Dertouzos, «On data banks and privacy homomorphisms,» Foundations of Secure Computation, pp. 169-180, 1978.

[2] C. Gentry, «A fully homomorphic encryption scheme.,» PhD thesis, Stanford University, 2009. http://crypto.stanford.edu/craig-thesis

[3] N. Smart and F. Vercauteren, «Fully Homomorphic Encryption with Relatively Small Key and Ciphertext Sizes.,» Cryptology ePrint Archive, Report 2009/571, 2009. http://eprint.iacr.org/571.pdf

[4] M. van Dijk, C. Gentry, S. Halevi and V. Vaikuntanathan, «Fully homomorphic encryption over the integers,» Cryptology ePrint Archive, Report 2009/616, 2009. http://eprint.iacr.org/..

[5] G. Chunsheng., «Fully Homomorphic Encryption Based on Approximate Matrix GCD,» Aavailable at eprint.iacr.org/2011/645.pdf.

[6] V. Vikuntanathan and Z. Brakerski, «Efficient Fully Homomorphic Encryption from (Standard) LWE,» Available at http://eprint.iacr.org/2011/344.pdf

[7] A. EL-YAHYAOUI et M. EL KETTANI, «Fully homomorphic encryption: state of art and comparison,» Available at:
https://www.academia.edu/25106824/Fully_Homomorphic_Enc ryption_State_of_Art_and_Comparison.

[8] J. Coron, A. Mandal, D. Naccache, M. Tibouchi, "Fully Homomorphic Encryption over the Integers with Shorter Public Keys". Available at http://eprint.iacr.org/2011/441

[9] J. Coron T. Lepoint and M. Tibouchi. "Batch fully homomorphic encryption over the integers." 2012. Available at eprint.iacr.org/2013/36

[10] J. Li and L. Wang, "Noise-free Symmetric Fully Homomorphic Encryption based on noncommutative rings", Cryptology ePrint Archive, Report 2015/641.

[11] M. Yagisawa " Improved fully homomorphic encryption with composite number modulus" Cryptology ePrint Archive, Report 2016/50, 2016. http://eprint.iacr.org

[12] G. Kristian et S. Martin , «Can there be efficient and natural FHE?,» *https://eprint.iacr.org/2016/105.pdf*